\documentclass[twocolumn]{article}
\usepackage{icqproc,epsf,epsfig,amssymb,amsmath}

\def\veps{\varepsilon}

\begin{document}
\thispagestyle{empty}

\twocolumn[
\vspace*{30mm}
\begin{LARGE} 
\begin{center}
%
%
Electronic stabilization of amorphous and quasicrystalline metals: 
Importance of quantum correlations
%
%
\end{center}
\end{LARGE}
\begin{large}
\begin{center} 
%
%
Hans Kroha$^{\dag}$ 
%
%
\end{center}
\end{large}
\begin{footnotesize}
\begin{it}
\begin{center}
%
%
Institut f\"ur Theorie der Kondensierten Materie,
Universit\"at Karlsruhe,
Postfach 6980,
76128 Karlsruhe, Germany
%
%
\end{center}
\end{it}
\end{footnotesize}
\begin{footnotesize}
\begin{center}
%
%
12 October 1999
%
%
\end{center}
\end{footnotesize}
\vspace{4ex}
\begin{small}
\hrule\vspace{3ex}
\begin{minipage}{\textwidth}
{\bf Abstract}\vspace{2ex}\\
\hp 
%
%
Numerous experimental indications suggest that the Hume-Rothery 
mechanism plays an important role in stabilizing quasicrystalline  
and amorphous phases. However, the exponential damping of the conventional 
Friedel oscillations at the relevant, elevated temperatures $T$ poses 
a severe challenge to the HR stabilization.
In order to resolve this problem it is shown using a Feynman diagram
technique that quantum correlations in the electron sea, arising from the
interplay of Coulomb interaction and impurity scattering, 
can strongly enhance the Friedel oscillations in these systems even
at elevated temperature. The resulting corrections to the 
Friedel potential are in agreement with available experimental results on 
amorphous HR alloys. It is proposed
to include the  enhancement of the Friedel amplitude
derived in the present work into 
pseudopotentials through the local field factor.
%
%
\vspace{2.5ex}
\end{minipage}\vspace{3ex}
\hrule
\end{small}\vspace{6ex}
]

%
%
\section{Introduction}

\hp

The microscopic origin of the stability of quasicrystalline phases
continues to be an unresolved issue. While at high temperatures the
long-range order of quasicrystals and approximants may be favored
over crystalline order because of entropic effects \cite{joseph.97},
striking interrelations between the ionic and the electronic 
structures \cite{poon.92,haberkern.99} especially in icosahedral (i) 
quasicrystals seem to indicate that electronic stabilization plays an 
important role at short and intermediate length scales of
of several atomic spacings:
(1) The electron density of states (DOS) has a pronounced, structure-induced
pseudogap at the Fermi level $\veps _F$. (2) The 
position $k_p$ of the main ionic structural peak coincides with
twice the Fermi wave number, $2k_F \simeq k_p$ \cite{madel.99}. 
(3) In dependence of 
the composition from their constituents, quasicrystalline
phases are only stable in small regions where the condition  
$2k_F \simeq k_p$ is satisfied. (4) In the quasicrystalline phase
the electrical resistivity is substantially larger than that of each
of the elemental constituents. Similar coincidences
between electronic and ionic structures are observed in amorphous
noble-polyvalent alloys \cite{madel.99,haeuss.92}. 
These findings may be traced back to the common feature of these materials
that the ion system exhibits concentrical, shell-like density correlations,
where the spacing $a$ between neighboring shells coincides with the
Friedel wavelength, $\lambda _F \equiv \pi /k_F = 2\pi /k_p \equiv a$.
Therefore, it has been conjectured that 
the electronic Friedel oscillations around 
an arbitrary central ion give an important contribution to the pair 
potential. As a consequence, the total energy of the system should
be optimized by the ions effectively being bound in the minima of the 
Friedel potential and a concommittant pseudogap formation at $\veps _F$.
The importance of such a Hume-Rothery (HR) stabilization mechanism 
is supported by detailed theoretical studies at temperature $T=0$
both for amorphous \cite{hafner.90} and for quasicrystalline 
\cite{ashcroft.87,fuji.91,hafner.92,widom.98} systems.
Numerical simulations \cite{dmitrienko.95} show that quasicrystalline
structures can indeed be grown by the use of pair potentials
with appropriate repulsive (i.e. oscillatory) parts. 

However, at the relevant temperatures 
where quasicrystals or amorphous structures are stable 
($T\simeq 10^2\dots 10^3 K$), regular Friedel oscillations are 
exponentially damped, and the thermal ion energy is so large that
the experimentally observed short- to intermediate-range concentrical
ion density correlations cannot be understood on the basis of the
conventional Friedel oscillations alone.

In the present work 
it is shown that the interplay between Coulomb electron-electron
interaction and disorder can lead to a strong enhancement as well as to
a systematical phase shift of the Friedel oscillations even at finite
$T$. Both effects compare 
well with available experimental results on amorphous HR alloys, and support
the validity of the HR mechanism even at elevated temperatures. 

\hp

\section{Model and effective interaction}

\hp

We here discuss an effective model for the electron motion
in amorphous and quasicrystalline systems at $T>0$. The ionic density
correlations mentioned above constitute a scattering potential for the
electrons whose scattering T-matrix $t_{\vec k,\vec k'}$ is, by definition,
proportional to the static ion structure factor,  
$t_{\vec k,\vec k'} \propto S( |\vec k - \vec k'| )$, which in turn
is peaked at a momentum transfer  $q \equiv |\vec k - \vec k'| = k_p
\simeq 2k_F$ and thus leads to enhanced backscattering.
As has been shown \cite{mahan.90,kroha.90,kroha.95}, the latter
not only generates a pseudogap 
but at the same time leads to a substantial increase of the 
electron transport or density relaxation rate, $\tau ^{-1}$, over the 
quasiparticle decay rate, $\tau _{qp}^{-1}$. 
$\tau ^{-1}$ is related to the conductivity $\sigma$ by
$\sigma = ne^2\tau / m^*$.  
Thus, $\tau ^{-1} > \tau _{qp}^{-1}$ is a generic feature of the 
amorphous and the quasicrystalline state. 
In quasicrystals, when the conductivity is substantially reduced
below the Drude result, we may have $\tau ^{-1} \gg \tau _{qp}^{-1}$.

In addition, at the relevant, finite temperatures the phase coherence
of the electrons is lost on the length scale of the inelastic mean free
path. It follows that the electrons cannot probe the long-range
order in a quasicrystal. Rather, they experience an effective potential
made up of randomly placed, spatially extended scattering centers, 
each one characterized by the T-matrix $t_{\vec k,\vec k'}$ \cite{kroha.99}. 
In such a potential the electronic motion is diffusive instead of ballistic,
similar as in amorphous metals. When the electron coherence length is long
enough (low $T$), the motion may be subdiffusive with a diffusion
exponent $\beta < 1/2$ \cite{bellissard.98,piechon.96}. However, for the present
purpose the precise value of $\beta$ is unimportant, and we will assume
$\beta =1/2$ (classical diffusion) in the following.

Diffusion, as a dissipative process, is difficult to incorporate in an
{\it ab initio} calculation. 
Therefore, we will
choose a Feynman diagram technique, where diffusion arises
by averaging over all (quasi-)random configurations of the system. 
In a diffusive electron sea screening is inhibited, so that the
effective Coulomb interaction, $v_q^{eff} (z,Z)$, between electrons with  
complex frequencies $z$ and $z+Z$ acquires a long-range,   
retarded part \cite{altsh.79},
\begin{eqnarray}
v_q^{eff}(z,Z)={v_q \Gamma ^2(z,Z,q)\over \epsilon ^{RPA}(Z,q)} ,\quad
v_q={4\pi e^2 \over q^2},                 
\label{veff}
\end{eqnarray}
where $\epsilon ^{RPA}(Z,q)\! =\! 1\! +\! 2\pi i\ \sigma\ /
(Z\, \mbox{sgn}Z''+iq^2 D)$ is the disordered 
RPA dynamical dielectric function,
and the diffusion vertex, defined in
Fig. 1 a), is
\begin{eqnarray}
\Gamma (z,Z,q) = \left\{  \begin{array}{ll}
       {i/\tau  \ \mbox{sgn}Z'' \over  
       Z+iq^2D\ \mbox{sgn}Z'' }\hfill &z''(z+Z)''<0  \\
       1 & \mbox{otherwise.}
                          \end{array}
                 \right.                   
\label{vertex}
\end{eqnarray}
$D=1/3\  v_F^2\tau $  
and $''$ denote the diffusion constant
and the imaginary part, respectively, and $\beta =1/2$.  
The long-range nature of $v_q^{eff}$ is a consequence of the
hydrodynamic ($Z,q\rightarrow 0$) pole of $\Gamma$, Eq. (2). 
Since diffusion is a classical phenomenon, guaranteed
by particle number conservation, the form Eq. (1)
of the effective interaction persists at finite $T$. 

There are two experimental indications for the diffusion model
of electron transport to be valid in i-quasicrystals. 
First, note that the diffusion-enhanced effective
Coulomb interaction Eq. (1) implies the well-known 
$\sqrt{|E-\veps_F|}$ dependence of the DOS in the pseudogap 
of disordered systems \cite{altsh.79},
where the half-integer power is characteristic for diffusion.
The fact that a powerlaw dip in the DOS at $\veps_F$ with an 
exponent very close to 1/2 has been observed in 
i-quasicrystals by tunneling measurements \cite{davydov.96}
may be taken as an indication that the electron motion is indeed
diffusive in these systems, and that the effective interaction has
indeed the Alt'shuler-Aronov form Eqs. (1), (2).
Second, the diffusion model, based on a finite phase coherence length,
explains why the spikiness of the DOS, predicted for ideal quasicrystals
at $T=0$, has up to now not been observed experimentally \cite{stadnik.96}.
\begin{figure}
\centerline{\psfig{figure=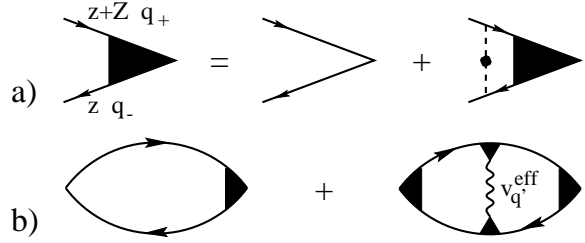,width=0.95\linewidth}}
\caption{a) Diagrammatic definition of the diffusion vertex $\Gamma$. 
b) Polarisation $\Pi (0,q)$
including leading order quantum correction induced by disorder and 
interactions. Dashed lines denote electron--ion scattering T-matrix  
$t_{\vec k,\vec k'}$,
the wavy line with solid triangles the effective Coulomb interaction.
\label{fig:polaris}}
\end{figure} 

\hp

\section{Electron density response} 

\hp

The incomplete screening of the electron-electron interaction in a
diffusive metal may be expected to drastically affect the screening charge
distribution around an ion in the electron sea as well. 
In order to calculate this effect on the 
Friedel oscillations, we must consider
the static charge density response $\chi (0,q)$ in the vicinity of $q=2k_F$.
It is given in terms of the polarization function $\Pi (0,q)$ as
$\chi (0,q)=\Pi (0,q)/(1-v_q\Pi (0,q))$. 
The first term of Fig. 1 b), $\Pi ^{(0)}(0,q)$,
corresponds to the well-known Lindhard function (RPA) \cite{mahan.90}. 
In this diagram
$\Gamma $ contributes only a nonsingular factor of $O(1)$, 
since here the effective interaction $v_q^{eff}$ enters in the
static limit at wave numbers $q \simeq 2k_F$, where the diffusion vertex
$\Gamma$ is structureless. We are thus led to consider quantum corrections where
$\Gamma$ gives contributions with vanishing frequency and momentum transfer,
so that the {\it hydrodynamic} transport properties become important,
although the response is taken at large external wave numbers $q$.
The most singular contribution of this type arises 
from the quantum correction $\Pi ^{(1)}(0,q)$, shown in Fig. 1 b), 
2nd diagram. 

While diffusive density relaxation occurs in general for large times, $t>\tau$,
in $\Pi ^{(1)}(0,q)$ it is, in addition, cut off for times 
larger than the life time $\tau _{qp}$ of the quasiparticles, which are
interacting via $v_{q'}^{eff}$. Thus, the frequency transfer in this term 
extends over the nonvanishing (see above) 
range $\tau _{qp}^{-1} \leq |Z| \leq \tau ^{-1}$. 
It may be evaluated explicitly as \cite{kroha.99,kroha.95},
\begin{eqnarray}
\vspace*{-1em}\Pi ^{(1)}(0,q) = - C \frac{2m^{*}k_F}{(2\pi \hbar )^2}
\frac{\tau _{qp}/\tau -1}{(\veps _F\tau )^{7/2}}\times \\
\int _{-\veps_F}^{\veps_F} \mbox{d}\nu  
{1/(4T) \over \mbox{cosh}^2{\nu\over 2T}}
\frac{x-1}{[(x-1)^2+(\frac{1}{4\sqrt{2}\veps_F\tau _{qp}})^2]^{3/4}}\ , \nonumber
\label{Pi3}
\end{eqnarray}
where $x=x(\nu )=(q/ 2k_F) / \sqrt{1+\nu / \veps _F}$ and 
$C$ is a numerical constant of $O(1)$. 
It is seen that for $T=0$, $\tau _{qp}^{-1} \rightarrow 0$
$\Pi ^{(1)}(0,q)$ exhibits a powerlaw divergence 
$\propto -\mbox{sgn}(q-2k_F)/|q-2k_F|^{1/2}$ 
at $q=2k_F$.
Although at finite $T$ or $\tau _{qp}^{-1}$ the divergence 
of $\Pi ^{(1)}(0,q)$ is reduced to a peak, the inverse dielectric 
function, $1/\veps (q) =1/(1-v_q \Pi (0,q))$,
still has a $q=2k_F$ divergence at a 
critical transport rate, $\tau _{c}^{-1}(T)$ because of the vanishing
denominator. This leads to a systematical 
enhancement as well as to a phase shift of the Friedel oscillations
(see below).
On the other hand, when there is no enhanced backscattering, we have
$\tau _{qp}/\tau = 1$, and the peak structure of $\Pi (0,q)$ vanishes.
Eq. (3) constitutes an 
extension of previous work \cite{kroha.95} in that the finite
quasiparticle life time is explicitly taken into account.
The parameter $\tau ^{-1}$ may be varied by changing the 
composition of the alloy.

\hp

\section{Comparison with experiments}

\hp

In the following, the theory developed in the previous section 
is applied in the region $1/\veps_F \tau _{qp} \ll 1$
to a large class of noble-polyvalent metal alloys like Cu$_{1-x}$Sn$_x$.  
These HR alloys exhibit an amorphous to crystalline 
transformation (CAT) as a function of the composition of the alloy,
and the thermal stability may be continuously varied.
Remarkably, in all these systems 
(1) the thermal stability reaches a maximum at or near the CAT (Fig. 2)
\cite{haeuss.92}, and (2), assuming that the ions sit in the minima of
the Friedel potential, the measured ionic positions suggest that there is
a systematical phase shift $\varphi$ of the Friedel oscillations 
\cite{haeuss.92},
$\rho (r) \propto cos (2k_F r -\varphi)$, with
$\varphi = \pi /2$ at the CAT (inset of Fig. 2).

\begin{figure}
\centerline{\psfig{figure=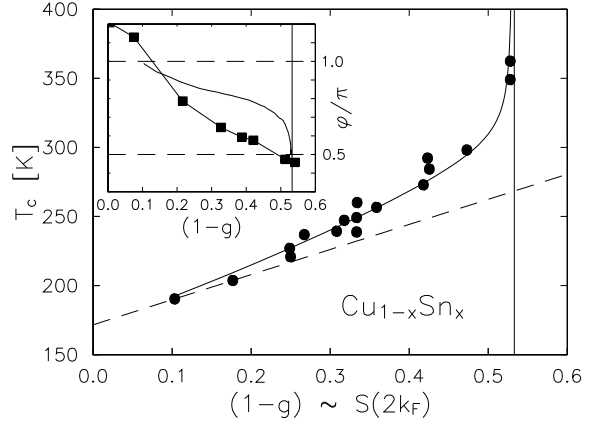,width=0.95\linewidth}}
\caption{Crystallization temperature $T_{c}$
as a function of the DOS suppression at $\veps _F$, $(1-g)$. 
Data points represent $T_{c}$ for a-$Cu_{1-x}Sn_x$ [5]. 
The solid curve is the fit of the present theory 
(see text). Vertical line: position of CAT.
The inset shows the phase shift $\varphi$ of the first maximum
of the charge density distribution $\rho (r)$.
Solid line: theory. Data points with solid line: measurements [5] for
a-$Cu_{1-x}Sn_x$.
\label{fig:Tcryst}}
\end{figure}

Fourier transforming $1-1/\veps (q)$ to obtain $\rho (r)$ \cite{mahan.90} 
shows that for incomplete Fermi surface-Jones zone matching, i.e.
small $\tau ^{-1}\simeq \tau ^{-1}_{qp}$, the quantum corrections 
generate density oscillations
$\rho ^{(1)}(r) \propto - \mbox{cos}(2k_Fr)/r^{3}$, which overcompensate 
the conventional  Friedel oscillations, 
implying a phase shift of $\varphi = \pi$ 
\cite{kroha.95}. As
$\tau ^{-1} \rightarrow \tau _{c}^{-1}$, the increasing $2k_F$
peak of $1/\veps (q)$ leads, in addition, to density oscillations   
$\rho ^{(1)}(r) \propto \mbox{sin}(2k_Fr)/r^{2}$, so that in the
vicinity of $\tau _{c}^{-1}$
\begin{eqnarray}
\rho (r)\propto -{\mbox{cos}(2k_Fr)\over (2k_Fr)^{3}} + 
         A(\tau ^{-1}){\mbox{sin}(2k_Fr)\over (2k_Fr)^{2}},
\label{rho}
\end{eqnarray}
with $A(\tau ^{-1})\simeq 0.343\pi (1-\tau ^{-1}/\tau _{c}^{-1})^{-1/2}$.
The exponent $1/2$ is characteristic for diffusive behavior.
Thus, the Friedel oscillations 
are shifted by $\varphi =\pi -\mbox{tan}^{-1}[2k_Fr~A(\tau ^{-1})]
\simeq \pi/2 + 1/(2k_FrA)$, i.e. the diverging
Friedel amplitude necessarily goes hand in hand 
with $\varphi \rightarrow \pi/2 +0$.
Note that, in contrast to the conventional Friedel oscillations, 
this divergence is robust against damping due to finite $T$ or disorder.
The point where the amplitude, $A$, diverges should be identified with the CAT,
since at this point the fluctuations of the Friedel potential also diverge,
allowing the system to find its crystalline ground state.
This process explains in a natural way 
the observed composition dependence of the thermal stability
and of the phase shift $\varphi$ mentioned at the beginning of this section. 

For a direct comparison with experiments the control parameter of the
theory, $\tau ^{-1}$, must be translated into a parameter
which is experimentally accessible: 
It follows from the scattering theory \cite{kroha.90,kroha.95} that 
$\tau ^{-1} = \tau _o^{-1} + \gamma ~S(2k_F)$, where the peak of the ionic
structure factor, $S(q=2k_F)$, controls the backscattering amplitude, 
$\gamma $ is a constant, and $\tau_o^{-1}$ is an offset due to 
momentum independent scattering. $S(2k_F)$ in turn is 
proportional \cite{haeuss.92} to
the measured, structure--induced suppression of the DOS $N(\veps _F)$ 
at the Fermi level, $1- N(\veps _F)/N_o(\veps _F)\equiv 1-g$, 
compared to the free electron gas, $N_o(\veps _F)$.  
The resulting fit of the crystallization temperature, $T_{c}$, is shown in 
Fig. 2, where the contribution to the stability coming from
the pseudogap formation is assumed to be linear in $(1-g)$ (dashed line).
The characteristic increase of $T_c$ at the CAT, explained by the present
theory, is clearly seen. 
The inset shows the calculated phase shift, $\varphi$, 
and the measured shift of the atomic nearest neighbor position
relative to the position of the 
first conventional Friedel minimum, $a_o = \pi / k_F$.
Note that there is no adjustable parameter in $\varphi$, once the fit
of $T_c$ has been performed. 
The overall behavior of the shift
is well explained by the theory; however, the experimental data approaches
$\varphi = \pi/2$ faster than predicted. The latter may be attributed to the
fact that, as seen from the discussion after Eq. (\ref{rho}), the 
higher--order Friedel minima approach $\varphi = \pi /2$ faster than the
first one. In this light, the agreement between
theory and experiment is remarkably good.
 
\hp

\section{Conclusion}

\hp 

The structural similarities \cite{poon.92} 
between amorphous alloys and {\it i}--quasicrystals
suggest that the quantum effect discussed above may be important 
in the latter systems as well. In fact, quasicrystals seem to 
fulfill all the necessary conditions for this effect to
occur, i.e. effectively diffusive electron 
motion \cite{bellissard.98,piechon.96}
and $\tau ^{-1} \gg \tau _{qp}^{-1}$. The latter is 
supported by the Fermi surface matching, i.e. by the experimental
observation \cite{davydov.96} and theoretical prediction 
\cite{ashcroft.87,fuji.91,hafner.92} 
of structure-induced pseudogaps. Moreover, another
more commonly known effect of disorder--enhanced Coulomb interaction, 
the $\sqrt{|E-\veps _F|}$ dependence of the DOS in the 
pseudogap \cite{altsh.79}, 
may have been already observed in {\it i}-quasicrystals by 
tunneling measurements of the DOS \cite{davydov.96}.  

In conclusion, we have shown that the Friedel oscillations are 
enhanced by Coulomb interaction in the presence of disorder and 
enhanced backscattering.
The results are relevant for the stability of amorphous and quasicrystalline
metals. It is proposed to include the enhanced Friedel potential 
calculated in the present work in the pseudopotential of more
quantitative {\it ab initio} calculations.	 

Numerous discussions with A.~G.~Aronov, P.~H\"aussler, R. Haberkern,
A.~Huck, T.~Kopp, and P.~W\"olfle are gratefully 
acknowledged. This work is supported by DFG through SP Quasikristalle.

\hp

\begin{footnotesize}
\begin{frenchspacing}

\end{frenchspacing}
\end{footnotesize}

\end{document}